.

# Sign-changing non-monotonic voltage gain of HfO$_2$/Parylene-C/SrTiO$_3$ field-effect transistor due to percolative insulator to two-dimensional metal transition

Alejandro Schulman, Ai Kitoh, Pablo Stoliar, and Isao H. Inoue

# Sign-changing non-monotonic voltage gain of HfO₂/Parylene-C/SrTiO₃ field-effect transistor due to percolative insulator to two-dimensional metal transition

Alejandro Schulman,[1] Ai Kitoh,[1] Pablo Stoliar,[1,2,a)] and Isao H. Inoue[1,a)]

[1]National Institute of Advanced Industrial Science and Technology (AIST), AIST Tsukuba Central 5, Tsukuba 305-8565, Japan
[2]CIC nanoGUNE, Tolosa Hiribidea 76, 20018 Donostia-San Sebastian, Spain

Controlling the insulator-to-2D-metal transition is a promising key to overcome the scaling problem that silicon-based electronic devices will face in the near future. In this context, we examine the channel formation of SrTiO₃-based solid-gated field-effect devices in which a 2D metal phase coexists with a semiconductor phase. A non-monotonic voltage-gain transfer characteristic with negative and positive slope regions is observed. We introduce a numerical model that helps to rationalize the experimental findings in terms of the established physics of field-effect transistors and percolation. Our numerical study not only reproduces the experimental results but also provides non-trivial predictions, which we verify experimentally. *Published by AIP Publishing.*
[http://dx.doi.org/10.1063/1.4973739]

Electrostatic control of carrier density is a key technique used in semiconductor devices. Nevertheless, the actual number of carriers in cutting-edge devices is approaching the detection limit as devices approach the nanoscale. Alternative ideas are needed to enable further miniaturization;[1] in particular, the use of a quasi-2D electron gas (2DEG) for the channel is a promising solution because the metal has a larger number of carriers even in nanoscale devices.

SrTiO₃ (STO) is a band insulator (band gap ∼3.2 eV)[2,3] with a threshold of metallicity $(8.5 \times 10^{15} \, cm^{-3})$[4] that features orders of magnitude lower than those of Si $(3.5 \times 10^{18} \, cm^{-3})$ and Ge $(3.5 \times 10^{17} \, cm^{-3})$. This low threshold of metallicity is believed to stem from the unique quantum paraelectricity and/or large and nonlinear dielectric response of STO.[5,6] This property is observed not only in the bulk but also on the surface, where a 2DEG[7–9] is present and the insulator-to-2D metal transition (IMT) occurs.[10–15] The charge confinement induces intriguing electronic properties, such as Rashba spin–orbit coupling due to inversion symmetry breaking.[10,16,17] The use of a 2DEG with a large carrier density is the key to solving the scaling problem. Therefore, STO is an archetypal material in the oxide electronics field.[18,19] However, the significant problem associated with STO is that the surface is very prone to oxygen-defect formation.

A recent advance in bilayer gate technology using Parylene-C has overcome this issue and enabled the development of a practical field-effect transistor (FET) with an STO channel.[20] The STO FET exhibits steep-slope (170 mV/decade) threshold behavior and a large carrier mobility (11 cm²/Vs) at room temperature.[14] This breakthrough enables the demonstration of a gate-induced IMT within the channel of an STO-based FET, which is a main result of this letter.

It is important to note that the bilayer-gate STO FETs studied thus far exhibit peculiar electrical characteristics that are thought to stem from the phase separation, i.e., a coexistence of phases with different electronic properties.[14] Such nontrivial electrical characteristics, which are qualitatively different from those of standard field-effect devices, should be further examined in a detailed study of the channel formation.

To address this problem, we performed a combined theoretical/experimental study. We identified the key features of an STO FET and developed a model to rationalize these features by assuming a set of minimal physical mechanisms. We were thus able to clarify the dynamics of the channel formation of an STO FET based on IMT.[14] We demonstrate that the dynamics provide STO FETs with a unique feature: controllable negative/positive voltage-gain transfer characteristic.

We fabricated the STO-based planar FETs (channel length $L = 20 \, \mu m$, width $W = 4 \, \mu m$) with extra potential-sensing electrodes, as shown in Fig. 1(a). First, a 3-nm-thick Parylene-C layer was deposited on highly insulating STO (001) substrates (Shinkosha Co., Ltd.) at room temperature using the Gorham method.[21] This Parylene-C layer both protected the STO surface during fabrication and served as part of the double-layer gate insulator. To deposit the source/drain electrodes directly on the STO, the Parylene-C film was selectively etched by ozone plasma under ultraviolet radiation. We then deposited 10-nm-thick Ti electrodes using e-beam evaporation (0.1 Å/s). Next, a second 3-nm-thick Parylene-C layer was deposited to cover the etched area uncovered by the Ti metal deposition. To complete the gate dielectric double layer, a 20-nm-thick HfO₂ layer was deposited on top of the Parylene-C layer using atomic layer deposition. Before the gate metal layer deposition, we deposited a 350-nm-thick SiO₂ layer in the area where the gate electrode pad was to be placed (not in the channel area). This additional insulating layer prevents body currents and resulted in a marked decrease of the gate leak current compared with devices

a)Authors to whom correspondence should be addressed. Electronic addresses: p.stoliar@aist.go.jp and isaocaius@gmail.com

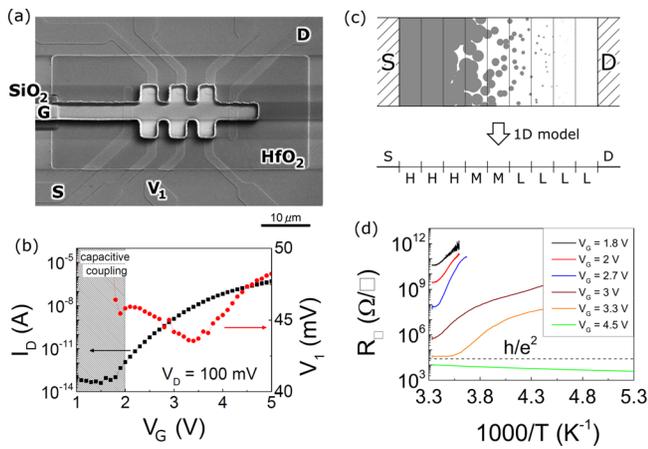

FIG. 1. (a) SEM image of the STO FET. In addition to the drain (D) and source (S) electrodes, extra voltage-probe electrodes are observed. The SiO$_2$ layer below the gate (G) pad is not part of the gate double layer but was added to avoid body currents. (b) Typical transconductance transfer characteristic ($I_D$ vs. $V_G$) and voltage gain transfer curve ($V_1$ vs. $V_G$). At low $V_G$, capacitive coupling dominates the signal in the $V_1$ electrode. (c) Rationale behind the procedure for mapping the 2D channel geometry into a 1D model. The 2D channel is divided into cells that, depending on local percolation, can exhibit high (H), medium (M), or low (L) conductance. (d) Arrhenius plot of the sheet resistance ($R_\square$ vs. 1/T) for different $V_G$ at $V_D = 0.1$ V. The quantum resistance of a 2DEG ($h/e^2 \approx 25.8$ kΩ) is indicated by the dashed line.

without the layer. The top gate electrode was deposited using radio-frequency sputtering of a 550-nm-thick Au layer with a 10-nm-thick Ti adhesion layer.

All the electrical measurements were performed using an Agilent Technology E5287A high-resolution source/monitor unit installed in an E5270B precision measurement mainframe. The room-temperature measurements were performed using a Karl Süss probe station. The low-temperature and magnetic field measurements were performed in a 9 T Quantum Design physical properties measurement system with a multi-function probe, where the sample was placed on a Kyocera PB-44713 ceramic quad flat package with ultrasonic bonded Al wires. The package was mounted on a custom-built IC socket fabricated by SDK Co., Ltd.

Figure 1(b) presents the transconductance transfer curve (drain current ($I_D$) vs. gate voltage ($V_G$)) at a constant drain voltage ($V_D$) = 0.1 V. The source and substrate were always grounded, and the gate leakage current was always below $10^{-10}$ A. The on/off ratio was greater than 7 orders of magnitude, with the off state equal to the noise level of the electrical instrumentation. The sub-threshold slope was 200 mV/dec. The linear-regime field-effect mobility was 20 cm$^2$/Vs ($V_G = 4$ V). These results qualitatively indicate that the STO surface remained unperturbed after the fabrication processes. The high quality of the STO surface resulted in sufficient reproducibility of the $I_D$–$V_G$ characteristics. The standard deviation of the onset voltage, where $I_D$ is clearly distinguished from noise, was 250 mV for more than 20 different devices on three substrates.

The multi-terminal configuration not only provides insight into transport mechanisms but also furnishes extra terminals of possible outputs. Here, we define an extra transfer characteristic in addition to the transconductance characteristic: the voltage at the midpoint of the channel $V_1$ as a function of $V_G$. Although the standard $I_D$ vs. $V_G$ curve is useful as a quality indicator, the $V_1$ vs. $V_G$ curve is also worth investigating, as described in this letter.

We focus on the $V_G > 2$ V region because capacitive coupling dominates the $V_1$ vs. $V_G$ curve at smaller $V_G$ (see supplementary material). A distinctive feature appears in the $V_1$ vs. $V_G$ curve (Fig. 1(b)) when $V_G$ is between 2.2 and 3.4 V, where $V_1$ decreases with increasing $V_G$. For higher $V_G$, $V_1$ monotonically increases without reaching saturation at $V_G = 5$ V. $V_1$ is expected to eventually approach a geometrical value of $V_D/2$. The non-monotonic characteristic of $V_1$ is associated with the formation of current filaments.[22] This negative slope region of the $V_1$ vs. $V_G$ curve is particularly interesting in certain applications, such as circuits using positive feedback (e.g., a Schmitt trigger or Colpitts oscillator) or requiring highly non-linear transfer curves (e.g., in neuromorphic engineering).

There is ample evidence regarding inhomogeneity in the 2DEG induced at the surface of STO, which has been studied under different conditions, e.g., electric double layer and LaXO$_3$/STO (X = Al,Ti) heterostructures.[23–28] Therefore, we applied the standard 1D model of FETs[29] to our device. This model assumes the gradual channel approximation. In this framework, the electric field generated by the gate only has a transverse component (perpendicular to the channel), which is only responsible for modulating the free carrier density in the channel. The longitudinal electric field generated by the channel bias, $V_D$, is instead only responsible for the carrier drift.[30] However, the filament formation or percolation in the inhomogeneous channel should be modeled in (at least) two dimensions. An effective approach to address the 2D parallel conducting paths in a 1D percolation model is to divide the channel into cells with areas $W \times dx$, where $x$ is the coordinate in the direction of the channel, and to evaluate the percolation in each cell individually. A schematic illustration of the model introduced in this work is presented in Fig. 1(c). The voltage drop at the element $dx$ is

$$dV_x = \frac{I_D}{W} G_x^{-1} dx,$$  (1)

where $V_x$ is the voltage profile in the channel and $G_x = e \mu n_x$ is the local sheet conductance ($e$ is the elementary charge, $\mu$ is the carrier mobility, and $n_x$ is the gate-induced local sheet carrier density).[29] In standard field-effect devices, the definition of $G_x$ is clear; $\mu$ is expected to be constant;[31] and the carrier density is defined by $n_x = \frac{C}{e} (V_G - V_x - V_{th})$, where $C$ is the total capacitance per unit area and $V_{th}$ includes interfacial potentials and the potential due to the non-mobile (trapped) charges. Because our FET is $n$-type and it is difficult to induce holes via the field effect, we can safely state that $n_x = \frac{C}{e} (V_G - V_x - V_{th})$ is zero whenever $(V_G - V_x - V_{th}) < 0$.

We previously reported the inhomogeneous nature of the induced channel in STO FET devices.[14,20] We proposed that an inhomogeneous energy landscape results in the coexistence of metallic and insulator patches. The ratio between these two patches is modulated by $n_x$ set by the local potential $(V_G - V_x)$. As $n_x$ increases, the metallic patches grow and eventually percolate at the percolation threshold $V_C$. Because $V_C$ and $V_{th}$ have different values, our devices exhibit two thresholds, which indicates the value of $V_G$ necessary to activate each phase (semiconductor-like or metallic-like). The cells become metallic when $(V_G - V_x) = V_C$. To address this issue, we introduce a non-monotonic scaling factor $\eta_x$

$$G_x = \eta_x \, \mu \, C \cdot (V_G - V_x - V_{th}). \qquad (2)$$

The physical description of $\eta_x$ is based on the percolation model; however, we avoid using the standard percolation formula because this is not standard percolation. The difference here is that the conductivity of the semiconductor matrix is non-zero. We thus use a generalized formula based on the error function:

$$\eta_x = \eta_0 + \eta_1 \operatorname{erf}\left(\frac{V_G - V_x - V_C}{s}\right), \qquad (3)$$

where $s$ is (non-trivially) related to the percolation critical exponent. $\eta_0$ and $\eta_1$ are nonessential scaling parameters; the only constraint is $\eta_x > 0$.

The criteria to set the model parameters are as follows. The region of the $V_I$ vs. $V_G$ curve with the "well" shape is adjusted by three parameters: $V_C$, $s$, and $\eta_1$. We estimated $V_C$ from the temperature measurements presented in Fig. 1(d), which show that the system becomes metallic at $V_G$ between 3.3 and 4.5 V. This transition has already been studied at low temperatures with a Parylene-C gate insulator.[17] However, using the bilayer gate insulator, we observe the IMT with $V_G$ values compatible with modern Si-based technology. Moreover, a good estimation of $V_G$ when the entire channel became metallic is provided. In our model, this behavior occurs at $V_G = V_C + V_D$ and not $V_G = V_C$ because the entire channel should be $V_G - V_x > V_C$. Because the midpoint between 3.3 and 4.5 V is 3.9 V (from Fig. 1(d)), we take $V_C = 3.8$ V ($V_D = 0.1$ V).

We set $s = 0.75$, which fixes the width of this region at values close to the experimentally observed values. We observed that the value should be between 0.1 (which results in a very abrupt transient) and 1 (where the negative slope region practically vanishes). $\eta_1$ controls the depth of this "well." We set this value to achieve a ten-fold enhancement in Eq. (3), i.e., $\frac{\eta_x(V_G \to +\infty)}{\eta_x(V_G \to -\infty)} = 10$, which ensures a qualitatively good match between the simulated and experimental data. $C$, $W$, $\mu$, and $\eta_0$ are linear dependent and can be absorbed by a factor set to have an approximate match between experimental and numerical $I_D$ values. Finally, $V_{th} = 1.9$ V was determined from statistics for many devices.

We compute the channel voltage profile and drain current (Fig. 2(a)) by integrating Eq. (1) with boundary conditions $V_x(0) = 0$ and $V_x(L) = V_D$, with $L$ representing the distance between the drain and source. The numerical results reported here were computed with actual experimental parameters (see supplementary material). The model reproduces the distinctive feature of the devices, that is, the non-monotonic voltage transfer curves.

The starting point from which to analyze the device behavior is the evolution of the carrier density profile with gate bias, shown in Fig. 2(b). The contour line corresponds to the carrier density that triggers the IMT, $n(V_G - V_x = V_C = 3.8$ V). Only the channel near the source electrode becomes metallic for $V_G = V_C$ because of the channel bias. Metallic patches then grow through the channel, which becomes fully metallic when $V_G > V_C + V_D$.

Figure 2(c) presents the derivative of the $V_I$ vs. $V_G$ curve for every position in the channel. The channel behavior, especially in the region where $\frac{dV_x}{dV_G}$ becomes negative, is visible. We superimposed the transition line from Fig. 2(b). The effect of the IMT begins before the transition line because of the broad slope of $\eta_x$ ($s = 0.75$ for Eq. (3)); the actual contribution of Eq. (3) starts at ~0.75 V, less than $V_C$ of the transition line. The $\frac{dV_x}{dV_G}$ value consequently increases for other regions of $V_G$, and when approaching $V_G = 5$ V, this value becomes approximately constant for all $x$ because the entire channel is metallic.

We used $V_D$ as a control parameter to corroborate our model. We performed the same simulation as that performed above but for different $V_D$ values and crosschecked the results with experimental data. As $V_D$ increases, the negative-gain region shifts toward higher $V_G$. Moreover, the transition region widens because $V_D$ becomes comparable to $V_G$ (see supplementary material).

In Fig. 3(a), we show the evolution of the minimum in $V_I$ as a function of $V_D$ for both the experimental and simulation data. The good agreement suggests that our model is valid for all $V_D$ in the linear operation regime of the STO FET.

Finally, we confirm the validity of our assumption: a strong discontinuity introduced by $\eta_x$. For this purpose, we measured the Hall coefficient of the carriers of our STO FET and deduced the sheet carrier density $n_\square$ as a function of $V_G$. The ratio between $n_\square$ and the geometrical sheet carrier density, $n_{geo} = \frac{c}{e}(V_G - V_x - V_{th})$, is $\eta_x = n/n_{geo}$ (for additional details, see supplementary material); we were able to obtain the absolute value of $\eta_x$ experimentally. To compare this experimental $\eta_x$ with the simulated value, we matched the value of $\eta_0$ with the experimental data. In Fig. 3(b), we present both the experimental and simulated $\eta_x$ values. $\eta_x$ ranges between 1 and 10, indicating an enhancement of $n_\square$. This enhancement was reported in our previous work.[14] Although an enhancement of $n_\square$ is compatible with the simulations, the discontinuity in $n_\square$ alone is sufficient to reproduce the transfer characteristics; i.e., the transfer characteristics can be well simulated for $\eta_x$ ranging between 0.1 and 1. Detailed studies are in progress to explain the $n_\square$ enhancement in the metallic phase.

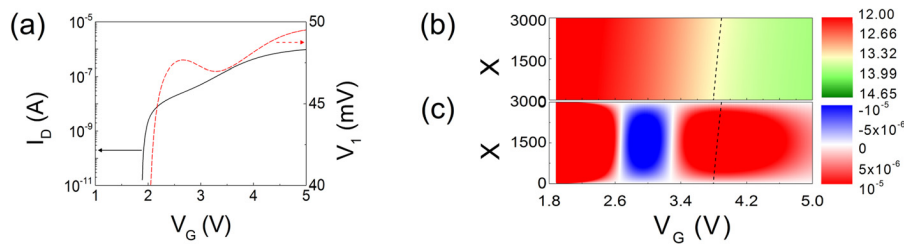

FIG. 2. (a) Simulated transconductance transfer characteristic and voltage gain transfer curve. (b) Computed contour plot of the local sheet carrier density. $X$ corresponds to the position in the channel; the source is located at $X = 0$. The color scale represents the logarithm of the local sheet carrier density. The dashed line corresponds to the local sheet carrier density required to trigger the IMT. (c) Computed contour plot of the derivative of the local potential with respect to the gate voltage, $dV_x/dV_G$. The z-scale (arbitrary units) presents negative-slope regions in blue and positive-slope regions in red. The dashed line was transferred from (b).

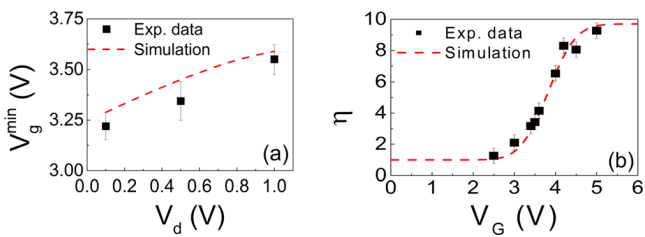

FIG. 3. (a) Shift in the position of the local minimum of the voltage gain transfer curve vs. $V_D$. The experimental points were obtained from statistical analysis of more than 10 devices. (b) Agreement between the predicted dependence of $\eta_x$ and the extracted one from the sheet carrier density obtained from Hall measurements.

Note that the parameters of Eq. (3) were set based on the temperature dependence of the sheet resistance for several $V_G$ values (to set $V_C$) and the transfer characteristics (to set $\eta_1$ and $s$), whereas the experimental data in Fig. 3(b) were obtained independently through Hall measurements, thus reaffirming the agreement between the simulated and experimental results.

In conclusion, after the addition of the second conduction mechanism to a 1D FET model, we were able to rationalize the nontrivial transfer characteristics, i.e., a non-monotonic voltage-gain transfer characteristic with negative and positive slope regions, of highly engineered STO FETs. Other physical properties such as the enhancement of the sheet carrier density and the effect of $V_D$ on the transfer characteristics were also studied. An understanding of these properties was essential to shed light on the channel transport physics. In particular, we observed that the system has two thresholds, $V_{TH}$ and $V_C$, associated with two conduction mechanisms. At low $V_G$ (between $V_{TH}$ and $V_C$), the channel behaves as a semiconductor, whereas at higher $V_G$ (above $V_C$), the conduction mechanism is dominated by the 2DEG metal. Percolation (filamentary path formation) occurs in between. This study is a step toward making use of the distinctive features of our STO FET in practical applications.

See supplementary material for details regarding the capacitive coupling, numerical methods, Hall effect measurements, and the use of $V_D$ as a control parameter.

We are grateful to N. Kumar and A. Sawa for valuable discussions and suggestions. P.S. acknowledges the support of the Spanish Ministry of Economy through the Ramón y Cajal (RYC-2012-01031). P.S. acknowledges the Japan Society for the Promotion of Science (JSPS) for the Invitation Fellowship for Research in Japan, and A.S. acknowledges the Postdoctoral Fellowship for Overseas Researchers. This study was supported by JSPS KAKENHI Grant Nos. 15H02113 (category A) and 15F15315 and by the National Institute for Materials Science (NIMS) as the "Nanotechnology Platform" Program of the Ministry of Education, Culture, Sports, Science and Technology (MEXT) in Japan.